\global\def\draftcontrol{0}

   \def\versionno{radiative corrections }
\catcode`\@=11

\expandafter\ifx\csname draftcontrol\endcsname\relax\global\def\draftcontrol{0}
\fi

{\count255=\time\divide\count255 by 60
\xdef\hourmin{\number\count255}
\multiply\count255 by-60\advance\count255 by\time
\xdef\hourmin{\hourmin:\ifnum\count255<10 0\fi\the\count255}}
\def\draftdate{\number\month/\number\day/\number\year\ \ \ \hourmin }

\newcommand\makepapertitle{\par
  \begingroup
    \renewcommand\thefootnote{\@fnsymbol\c@footnote}%
    \def\@makefnmark{\rlap{\@textsuperscript{\normalfont\@thefnmark}}}%
    \long\def\@makefntext##1{\parindent 1em\noindent
            \hb@xt@1.8em{%
                \hss\@textsuperscript{\normalfont\@thefnmark}}##1}%
     \newpage
     \global\@topnum\z@   % Prevents figures from going at top of page.
     \@makepapertitle
     \thispagestyle{empty}\@thanks
  \endgroup
  \setcounter{footnote}{0}%
  \global\let\thanks\relax
  \global\let\makepapertitle\relax
  \global\let\@makepapertitle\relax
  \global\let\@thanks\@empty
  \global\let\@author\@empty
  \global\let\@date\@empty
  \global\let\@title\@empty
  \global\let\title\relax
  \global\let\author\relax
  \global\let\date\relax
  \global\let\and\relax
  \def\version{\let\version\@version\@gobble}
}
\def\@makepapertitle{%
  \newpage
   \ifnum\draftcontrol=1 {}
   \version\versionno
   \vskip 3em%
   \else
   \hfill\hbox to 3cm {\parbox{4cm}{\@pubnum}\hss}%
   \vskip 3em%
   \fi
   \begin{center}%
   \let \footnote \thanks
     {\LARGE {\@title}}%
     \vskip 1.5em%
     {\normalsize%\large
       \lineskip .5em%
       \begin{tabular}[t]{c}%
         \@author
       \end{tabular}\par}%
     \vskip 1.5em%
     {\@bstract}%
     \end{center}%
     \vskip 1.5em
     \@date%
   \par
}

\gdef\@pubnum{}
\def\pubnum#1{%
  \gdef\@pubnum{#1}}

\gdef\@bstract{}
\def\Abstract#1{%
  \gdef\@bstract{%
   \parbox{\textwidth-0pc}{%
   \centerline{\bf Abstract}\penalty1000%
\kern.2cm%
\noindent%\abstractfont \baselineskip=12pt
\renewcommand\baselinestretch{1.0}%
{#1}}}
}

\def\ps@paper{\let\@mkboth\@gobbletwo%
     \ifnum\draftcontrol=1
    \def\@oddfoot{\hbox to \textwidth{\tiny \versionno \hfil\tiny\draftdate}%
    \hskip -\textwidth \hbox to \textwidth{\hfil\rm\thepage\hfil}}%
     \else\def\@oddfoot{\hbox to \textwidth{\hfil\rm\thepage\hfil}}
     \fi
     \let\@evenfoot\@oddfoot
}
\def\body{\clearpage
          \pagestyle{paper}
    }
\def\@version#1{\ifnum\draftcontrol=1
\typeout{}\typeout{#1}\typeout{}
\vskip3mm\centerline{\hbox{\fbox{\normalsize{\tt DRAFT -- #1 -- }
                   {\draftdate}}}}\vskip3mm
\fi}
\let\version\@version
\long\def\eqlabel#1{\ifnum\draftcontrol=1
                    \tag@false  % there are some problems with multline without this
                    \tag*{(\theequation) \hbox to -0.2cm{\hspace{0cm}\small{#1}\hss}}
                    \refstepcounter{equation}
                    \edef\@currentlabel{\theequation}
                    \ltx@label{#1}          % use old LaTeX \label instead of new definition
                    \else
                    \label{#1}
                    \fi
                    }
\let\st@bibitem\@bibitem
\let\st@lbibitem\@lbibitem
\ifnum\draftcontrol=1
  \def\@bibitem#1{%
    \st@bibitem{#1}\a@@label{#1}\ignorespaces}
  \def\@lbibitem[#1]#2{%
    \st@lbibitem[#1]{#2}\a@@label{#2}\ignorespaces}
  \def\a@@label#1{%
    \gdef\a@lab{\smash{\normalfont\small#1}}
    \ifvmode
      \if@inlabel
        \global\setbox\@labels\hbox{%
          \llap{\a@lab\let\a@lab\relax
                \kern\@totalleftmargin\kern\marginparsep}%
          \box\@labels}%
      \fi
    \fi}
\fi
\documentclass[12pt,letterpaper]{article}
\usepackage{amsmath,amssymb,array,calc,epsfig}
\usepackage{graphicx}
\usepackage{psfrag,verbatim,bm}
\ifnum\draftcontrol=1
\fi
\renewcommand\baselinestretch{1.25}
\renewcommand\section{\@startsection {section}{1}{\z@}%
                                   {-3.5ex \@plus -1ex \@minus -.2ex}%
                                   {2.3ex \@plus.2ex}%
                                   {\normalfont\large\bfseries}}
\renewcommand\subsection{\@startsection{subsection}{2}{\z@}%
                                   {-3.25ex\@plus -1ex \@minus -.2ex}%
                                   {1.5ex \@plus .2ex}%
                                   {\normalfont\normalsize\bfseries}}
\renewcommand\subsubsection{\@startsection{subsubsection}{3}{\z@}%
                                   {-3.25ex\@plus -1ex \@minus -.2ex}%
                                   {1.5ex \@plus .2ex}%
                                   {\normalfont\normalsize\it}}
\renewcommand\paragraph{\@startsection{paragraph}{4}{\z@}%
                                   {-3.25ex\@plus -1ex \@minus -.2ex}%
                                   {1.5ex \@plus .2ex}%
                                   {\normalfont\normalsize\bf}}

\numberwithin{equation}{section}

\def\ie{{\it i.e.}}

\def\revise#1       {\raisebox{-0em}{\rule{3pt}{1em}}%
                     \marginpar{\raisebox{.5em}{\vrule width3pt\
                     \vrule width0pt height 0pt depth0.5em
                     \hbox to 0cm{\hspace{0cm}{%
                     \parbox[t]{4em}{\raggedright\footnotesize{#1}}}\hss}}}}

\def\cali         {{\cal I}}

\def\call         {{\cal L}}

\def\calq         {{\cal Q}}

\def\del          {\partial}

\def\tr           {\mathop{\rm Tr}}

\def\sqr#1#2{{\vcenter{\vbox{\hrule height.#2pt
 \hbox{\vrule width.#2pt height#1pt \kern#1pt
 \vrule width.#2pt}\hrule height.#2pt}}}}

\def\a{\alpha}
\def\b{\beta}
\def\l{\lambda}

\def\dd{\delta}

\def\s{\sigma}
\def\e{\epsilon}
\def\g{\gamma}

\catcode`\@=12

\begin{document}

%%%
%%%%%% text starts here
%%%%%%%%%

\title{\bf Radiative Corrections in Vector-Tensor Models}
\pubnum
{UWO-TH-09/10
}

\date{May 2009}

\author{
A.~Buchel$ ^{1,2}$, F.~A.~Chishtie$ ^1$, M.~T.~Hanif$ ^1$, S.~Homayouni$ ^1$,\\ J.~Jia$ ^1$ and D.~G.~C.~McKeon$ ^{1,3}$\\[0.4cm]
\it $ ^1$Department of Applied Mathematics\\
\it University of Western Ontario\\
\it London, Ontario N6A 5B7, Canada\\
\it $ ^2$Perimeter Institute for Theoretical Physics\\
\it Waterloo, Ontario N2J 2W9, Canada\\
\it $ ^3$Department of Mathematics and Computer Science\\
\it Algoma University\\
\it Sault Ste. Marie, Ontario, Ontario P6A 2G4, Canada
}

\Abstract{
We consider a two-form antisymmetric tensor field $\phi$ minimally
coupled to a non-abelian vector field with a field strength
$F$. Canonical analysis suggests that a pseudoscalar mass term $\frac
{\mu^2}{2} \tr (\phi\wedge \phi)$ for the tensor field eliminates
degrees of freedom associated with this field. Explicit one loop
calculations show that an additional coupling $m\tr(\phi\wedge F)$
(which can be eliminated classically by a tensor field shift)
reintroduces tensor field degrees of freedom. We attribute this to the
lack of the renormalizability in our vector-tensor model. We also
explore a vector-tensor model with a tensor field scalar mass term
$\frac {\mu^2}{2} \tr (\phi\wedge\star \phi)$ and coupling
$m\tr(\phi\wedge \star F)$.  We comment on the Stueckelberg mechanism
for mass generation in the Abelian version of the latter model.
}

\makepapertitle

\body

\version\versionno

\section{Introduction}
We consider a model in which an antisymmetric tensor field $\phi_{\mu\nu}^a$ and a vector field $W_\l^a$ interact,
both through covariant derivatives and through direct coupling of $\phi_{\mu\nu}^a$ with the field strength $F_{\mu\nu}^a(W_\l^a)$. 
A pseudoscalar mass term for the tensor field has been shown to eliminate degrees of freedom associated with this field 
\cite{g1,g2}
and the consequences 
of this has been explored in the calculation of the vector and the tensor self-energy in \cite{g3,g4}. In section 2 we compute the 
the one-loop corrections to the mixed vector-tensor propagator and show that the direct coupling between 
$\phi_{\mu\nu}^a$ and $F_{\mu\nu}^a$ results in the breakdown of an identity derived in \cite{g4}. Thus, although tensor field
degrees of freedom can be eliminated classically, they must reappear at one-loop level. We attribute this to the lack of renormalizability 
in our model. 

In section 3 we discuss a slight variant of a vector-tensor model \cite{g3,g4}. Specifically, 
we replace the  tensor field pseudoscalar coupling and the mass term with the scalar ones. 
For a generic tensor field mass $\kappa$  and a coupling $m$ the theory is non-renormalizable.
When $\kappa^2+2 m^2=0$ the $U(1)$ version of the model is renormalizable. It generalizes the Stueckelberg model for 
a massive vector boson to that of a massive rank-two tensor field. We point out that the generalized Stueckelberg 
invariance leads to non-local transformations on the coupled matter fields.

\section{Anomalies in vector-tensor models with  pseudoscalar tensor field mass term and $\tr(\phi\wedge F)$ coupling}

The Lagrange density\footnote{We use mostly negative signature convention.}
\begin{equation}
\call=-\frac 14 F_{\mu\nu,a}F^{\mu\nu,a}+\frac{1}{12} G_{\mu\nu\l,a}G^{\mu\nu\l,a}+\frac m4 \e_{\mu\nu\l\s}\phi^{\mu\nu}_a F^{\l\s,a}
+\frac{\mu^2}{8}\e_{\mu\nu\l\s}\phi^{\mu\nu}_a\phi^{\l\s,a}
\eqlabel{defl}
\end{equation}
with 
\begin{equation}
\begin{split}
F_{\mu\nu}^a=&\del_\mu W_\nu^a-\del_\nu W_\mu^a+g f^{abc}W_{\mu, b}W_{\nu, c}\\
G_{\mu\nu\l}^a=&D_\mu^{ab}\ \phi_{\nu\l, b}+D_\nu^{ab}\ \phi_{\l\mu, b}+D_\l^{ab}\ \phi_{\mu\nu, b}\\
D_\mu^{ab}=&\dd^{ab}\del_\mu+g f^{acb} W_{\mu,c}
\end{split}
\eqlabel{der}
\end{equation}
for the vector field $W_\mu^a$ and the adjoint antisymmetric tensor field 
$\phi_{\mu\nu}^a$ was shown in \cite{g1} to have only two dynamical degrees of freedom
(those of the transverse polarization of the vector $W_\mu^a$) provided $\mu^2\ne0$. This is consistent with 
the results of \cite{g2} where the $m\to 0$, $g\to 0$, $U(1)$ limit of this Lagrange density was considered. 

The Euclidean space propagators for these fields appeared 
in \cite{g3},
\begin{equation}
\begin{split}
\langle W_\mu^a , W_\nu^b\rangle =&\frac{\dd_{\mu\nu}\dd^{ab}}{k^2}
\end{split}
\eqlabel{propa}
\end{equation}
\begin{equation}
\begin{split}
\langle \phi_{\a\b}^a,\phi_{\gamma\dd}^b\rangle=&\frac{\dd^{ab}}{\mu^4}\left(1+\frac{m^2}{k^2}\right)
\left(\dd_{\a\g} k_\b k_\dd-\dd_{\b\g}k_\a k_\dd+\dd_{\b\dd}k_\a k_\g-\dd_{\a\dd}k_\b k_\g\right)-\frac{\dd^{ab}}{\mu^2}\e_{\a\b\g\dd}
\end{split}
\eqlabel{propb}
\end{equation}
\begin{equation}
\begin{split}
\langle W_\mu^a,\phi_{\a\b}^b\rangle=&-\frac{i m \dd^{ab}}{\mu^2k^2}\left(\dd_{\a\mu}k_\b-\dd_{\b\mu}k_\a\right)
\end{split}
\eqlabel{propc}
\end{equation}
where the Feynman-'t Hooft gauge (with the gauge fixing term $\call_{gf}=-\frac 12 \left(\del_\a W^\a\right)^2$)
was used.  

The classical analysis  showed in 
\cite{g1,g2} that the tensor field has no dynamical degrees of freedom. 
Explicit computation \cite{g3} of the one loop radiative corrections 
to the two point function $\langle W_\mu^a,W_\nu^b\rangle$ leads to 
a cancellation of all contributions  from the 
diagrams containing the tensor field
(\ie, the pure gauge theory result is recovered). 

One can also see from \eqref{propb} that if $m=0$, then the tensor propagator 
has no poles. This is consistent with the tensor having no physical degrees 
of freedom when $\mu\ne 0$. It also implies that all radiative corrections 
vanish in the limit that there is no vector field even if some self 
interactions for the tensor field such as 
$(\phi_{\mu\nu}^a\phi_{a}^{\mu\nu})^2$ or $(\phi_{\mu\nu}^a\phi_{\lambda,a}^\nu
\phi_{\s}^{\l,b}\phi_{b}^{\mu\s})$ were present.
This is most easily seen when one uses dimensional regularization,
as in this case we only encounter integrals of the form 
$\int d^n k\ f(k)$ where $f$ is a polynomial function; such tadpoles 
are regulated to zero.

However, if $m\ne 0$ a pole does appear in the propagator of \eqref{propb}. 
This would appear to render the theory non-renormalizable as the asymptotic behavior 
of this propagator for large value of $k^2$ is that it grows as $k^2$. However, as pointed out 
in \cite{g4}, the shift 
\begin{equation}
\phi_{\mu\nu}^a=\chi_{\mu\nu}^a-\frac{m}{\mu^2} F_{\mu\nu}^a
\eqlabel{shift}
\end{equation}
in \eqref{defl} leads to (using $\e^{\mu\nu\l\s} D_{\nu,b}^a F^b_{\l\s}=0$)
\begin{equation}
\begin{split}
\call=&-\frac 14 F_{\mu\nu,a}F^{\mu\nu,a}+\frac{1}{12} H_{\mu\nu\l,a}H^{\mu\nu\l,a}
+\frac{\mu^2}{8}\e_{\mu\nu\l\s}\chi^{\mu\nu}_a\chi^{\l\s,a}+\frac {m^2}{8\mu^2} \e_{\mu\nu\l\s}F^{\mu\nu}_a F^{\l\s,a}
\end{split}
\eqlabel{lshift}
\end{equation}
where $H^a_{\mu\nu\l}$ is defined in terms of $\chi_{\mu\nu}^a$ and $W_\l^a$ in the same way that $G_{\mu\nu\l}^a$ is defined in terms of 
$\phi_{\mu\nu}^a$ and $W_\l^a$. Since the last term in \eqref{lshift} is topological, we see that it cannot contribute to a perturbative 
calculation; the coupling $m$ in \eqref{defl} has been removed and hence there is no pole in the propagator 
$\langle \chi_{\mu\nu}^a\chi_{\l\s}^b\rangle$.

The Feynman rules for the model defined by \eqref{lshift} are thus the same as those of the model of \eqref{defl} in the limit $m\to 0$. 
It is immediately apparent then that the one loop corrections to the two point functions $\langle \chi_{\mu\nu}^a\chi_{\l\s}^{b}\rangle$,
$\langle W_{\mu}^a\chi_{\l\s}^b\rangle$, $\langle \chi_{\mu\nu}^aW_{\l}^{b}\rangle$ all vanish.

By \eqref{shift}, we see then that 
\begin{equation}
\langle\phi_{\mu\nu}^a\phi_{\l\s}^b\rangle=\langle\chi_{\mu\nu}^a\chi_{\l\s}^b\rangle-
\frac{m}{\mu^2}\left(\langle F_{\mu\nu}^a\chi_{\l\s}^b\rangle+\langle\chi_{\mu\nu}^aF_{\l\s}^b\rangle\right)
+\frac{m^2}{\mu^4}\langle F_{\mu\nu}^aF_{\l\s}^b\rangle
\eqlabel{ppa}
\end{equation}
or
\begin{equation}
\langle\chi_{\mu\nu}^a\chi_{\l\s}^b\rangle=\langle\phi_{\mu\nu}^a\phi_{\l\s}^b\rangle+
\frac{m}{\mu^2}\left(\langle F_{\mu\nu}^a\phi_{\l\s}^b\rangle+\langle\phi_{\mu\nu}^aF_{\l\s}^b\rangle\right)
+\frac{m^2}{\mu^4}\langle F_{\mu\nu}^aF_{\l\s}^b\rangle
\eqlabel{ppb}
\end{equation}
In \eqref{ppa}, the only non-zero term on the right side of the equation is the last one; this leads to 
\begin{equation}
\langle\phi_{\mu\nu}^a\phi_{\l\s}^b\rangle=\frac{m^2}{\mu^4}\langle F_{\mu\nu}^aF_{\l\s}^b\rangle
\eqlabel{ppc} 
\end{equation}
Eq.~\eqref{ppb} then would result in 
\begin{equation}
\langle F_{\mu\nu}^a\phi_{\l\s}^b\rangle+\langle\phi_{\mu\nu}^aF_{\l\s}^b\rangle=-\frac{2m}{\mu^2} \langle F_{\mu\nu}^aF_{\l\s}^b\rangle
\eqlabel{ppd}
\end{equation}
However, the explicit computation of the 30 one loop diagrams contributing to $\langle \phi_{\mu\nu}^a\phi_{\l\s}^b\rangle$ 
in \eqref{ppc} leads to a pole contribution at $\e=2-\frac n2=0$ \cite{g4}
\begin{equation}
\begin{split}
\frac{m^2}{48p^2 \mu^8}\ \left(\frac{g^2 C_2 \dd^{ab}}{\pi^2\e}\right)\biggl[&(12p^4+5 p^2 m^2+14\mu^4)(p_\mu p_\l\dd_{\nu\s}
-p_\nu p_\l\dd_{\mu\s}+p_\nu p_\s\dd_{\mu\l}\\
&-p_\nu p_\s\dd_{\mu\l})-6\mu^2p^4\e_{\mu\nu\l\s}+3p^2\mu^4(\dd_{\mu\l}\dd_{\nu\s}-\dd_{\mu\s}\dd_{\nu\l})\biggr]
\end{split}
\eqlabel{pole}
\end{equation}  
Not only is this result inconsistent with renormalizability, but also it is clear that with $\langle F_{\mu\nu}^aF_{\l\s}^b\rangle$ 
being just the usual (transverse) Yang-Mills result, \eqref{ppc} and \eqref{pole} are incompatible.

This has motivated the computation of  the mixed propagator 
$\langle F_{\mu\nu}^a\phi_{\l\s}^b\rangle+\langle\phi_{\mu\nu}^aF_{\l\s}^b\rangle$ at one loop order. 
The two point contribution to these diagrams comes from $\langle\phi_{\a\b}^a,W_\l^b\rangle$ 
which has 30 diagrams contribution at one loop order. The sum contains the pole term
\begin{equation}
\begin{split}
\frac{19}{96\pi^2}\ \left(\frac{ iC_2 m g^2}{p^2 \mu^2\epsilon}\right)\biggl[p_\b \dd_{\a\l}-p_\a \dd_{\b\l}\biggr] \dd^{ab}
\end{split}
\eqlabel{mixedprop}
\end{equation}
This is not in accordance with \eqref{ppd}.

We are thus forced to conclude that the contact term $\frac m4 \phi_{\mu\nu}^aF^{\mu\nu,a}$ in \eqref{defl} destroys the renormalizability 
of the theory and thus the identities of \eqref{ppc} and \eqref{ppd} do not hold beyond the classical level. This is 
consistent with lack of renormalizability found when $m\ne 0$ in the model in which there are the interactions 
\[
g\ \bar{\psi}\g^\mu\tau^aW_{\mu,a}\psi+h\ \bar{\psi}\s^{\mu\nu}\tau^a\phi_{\mu\nu,a}\psi
\]
with a spinor field $\psi$ \cite{g3}.

We also note that if $m=0$ and a scalar mass term $\frac{\kappa^2}{2}\phi_{\mu\nu}^a\phi^{\mu\nu,a}$ is added to the Lagrangian 
of \eqref{defl}, then the propagator for $\phi_{\mu\nu}^a$ is inconsistent with renormalizability for all values of $\mu^2$.

\section{Vector-tensor models with tensor field  scalar mass and\\ $\tr (\phi\wedge F)$ coupling}

It is also worth considering a scalar mass for the tensor field and a direct scalar coupling between the tensor and 
the vector field strength. In this case we have \cite{kolb}
\begin{equation}
L=-\frac14 F_{\mu\nu}^aF^{\mu\nu}_a+\frac {1}{12}G_{\mu\nu\l}^aG^{\mu\nu\l}_a+\frac{\kappa^2}{2}\phi_{\mu\nu}^a\phi^{\mu\nu}_a
+m \phi_{\mu\nu}^aF^{\mu\nu}_a
\eqlabel{l3}
\end{equation}
Again, using the gauge fixing $L_{gf}=-\frac 12 \left(\del_\a W^\a\right)^2$, we find that the tensor propagator is 
\begin{equation}
-\frac{4\cali}{\del^2-2\kappa^2}+\frac{4(\del^2+4m^2)\calq}{\del^2(\del^2-2\kappa^2)(\kappa^2+2m^2)}
\eqlabel{tensorp}
\end{equation}
where 
\begin{equation}
\begin{split}
\cali_{\a\b\g\dd}=&\frac 12 \left(\dd_{\a\g}\dd_{\b\dd}-\dd_{\a\dd}\dd_{\b\g}\right)\,,\\
\calq_{\a\b\g\dd}=&\frac 14 \left(\dd_{\a\g}\del_\b\del_\dd-\dd_{\a\dd}\del_\b\del_\g+\dd_{\b\dd}\del_\a\del_\g
-\dd_{\b\g}\del_\a\del_\dd\right)
\end{split}
\eqlabel{def12}
\end{equation}
and the vector propagator is 
\begin{equation}
\frac{2\dd_{\mu\nu}}{\del^2}+\frac{4m^2(\del_\mu\del_\nu-\del^2\dd_{\mu\nu})}{\del^4(\kappa^2+2m^2)}
\eqlabel{vper}
\end{equation}
The second term in \eqref{tensorp} indicates that the model of \eqref{l3} is not renormalizable. However, 
we also note that if $\kappa^2+2m^2=0$ the propagators of \eqref{tensorp} and \eqref{vper} are ill defined.

If we set $\kappa^2=-2m^2$ in \eqref{l3}, $L$ reduces to 
\begin{equation}
L=\frac{1}{12}G_{\mu\nu\l}^aG^{\mu\nu\l}_a-m^2\left(\phi_{\mu\nu}^a-\frac{1}{2m}F_{\mu\nu}^a\right)^2
\eqlabel{lsp}
\end{equation}
In the $U(1)$ limit (which serves to pick out those terms in \eqref{lsp} that are bilinear in the fields)
\eqref{lsp} is invariant under the transformation 
\begin{equation}
\begin{split}
\phi_{\mu\nu}&\ \to\ \phi_{\mu\nu}+\del_\mu\Theta_\nu-\del_\nu\Theta_\mu\\
W_\mu&\ \to\ W_\mu+2m\Theta_\mu
\end{split}
\eqlabel{17}
\end{equation}
This gauge invariance is analogous to that of the Stueckelberg model for a massive $U(1)$ 
vector boson where \cite{stue}
\begin{equation}
L=-\frac 14 F_{\mu\nu}F^{\mu\nu}-\frac{m^2}{2}\left(W_\mu-\frac 1m\del_\mu\s\right)^2
\eqlabel{18}
\end{equation}
which is invariant under the transformations 
\begin{equation}
W_\mu\ \to\ W_\mu+\del_\mu\Theta
\eqlabel{19a}
\end{equation}
\begin{equation}
\s\ \to\ \s+m\Theta
\eqlabel{19b}
\end{equation}

The gauge fixing 
\begin{equation}
L_{gf}=-\frac{1}{2\a}\left(\del^\mu\phi_{\mu\nu}-2\a m W_\mu\right)^2
\eqlabel{lgf}
\end{equation}
when added to $L$ in \eqref{l3} serves to decouple the fields $\phi_{\mu\nu}$ and $W_\mu$. The propagators for both 
$\phi_{\mu\nu}$ and $W_\mu$ are consistent with renormalizability. However, coupling with matter fields must be 
consistent with the invariance of \eqref{17}. The vector field $W_\mu$ can be coupled to a matter field 
$\psi$ (a spinor or scalar) if we use covariant derivative 
\begin{equation}
D_\mu\psi=\left(\del_\mu+i\ W_\mu\right)\psi
\eqlabel{covd}
\end{equation}
provided $\psi$ undergoes the non-local transformation
\begin{equation}
\psi(x)\ \to\ \exp\left(-2i m\int_{x_0}^x dy^\mu \Theta_\mu(y)\right)\ \psi(x)
\eqlabel{psit}
\end{equation}
when $W_\mu$ undergoes the transformation of \eqref{17}. Note that if $\Theta_\mu=\frac{1}{2m}\del_\mu \omega$
\eqref{psit} reduces to the standard gauge transformation of a matter field. 
The propagator for the field $W_\mu$ is
\begin{equation}
\left[\left(\del^2+4\a m^2\right)\dd_{\mu\nu} -\del_\mu\del_\nu\right]^{-1}
=\frac{\dd_{\mu\nu}}{\del^2+4\a m^2} +\frac{\del_\mu\del_\nu}{4\a m^2(\del^2+4\a m^2)}
\eqlabel{23}
\end{equation}
which indicates the presence of a gauge dependent non zero pole in the propagator for 
$W_\mu$.

There does not appear to be a way of coupling either the tensor $\phi_{\mu\nu}$ of \eqref{lsp} 
or the Stueckelberg scalar $\s$  of \eqref{18} to matter that respect the gauge invariance of \eqref{17} and \eqref{19b} correspondingly.
However, the invariance of \eqref{17} and \eqref{19a} can be respected when the $U(1)$ vector 
field $W_\mu$ is coupled to matter; this possibility has been incorporated into the Standard Model 
\cite{kuzmin,g8}.

The invariance of \eqref{17} has no non-Abelian extension. However, $L$ in \eqref{l3}
can be replaced by
\begin{equation}
\begin{split}
L=&\frac{1}{12}\left[G_{\mu\nu\l}^a+g f^{abc}\left(F_{\mu\nu,b}\s_{\l,c}+
F_{\nu\l,b}\s_{\mu,c}+F_{\l\mu,b}\s_{\nu,c}\right)\right]^2
\\
&-m^2\left[\phi_{\mu\nu}^a+\left(D_{\mu}^{ab}\s_{\nu,b}-D_{\nu}^{ab}\s_{\mu,b}\right)\right]^2
\end{split}
\eqlabel{24}
\end{equation} 
where $G_{\mu\nu\l}^a$, $F_{\mu\nu}^a$ and $D_{\mu}^{ab}$ are as in \eqref{der}. Eq.~\eqref{24}
is invariant under the usual Yang-Mills gauge transformations as well as the transformations 
\begin{equation}
\begin{split}
\phi_{\mu\nu}^a&\ \to\ D_{\mu}^{ab}\Theta_{\nu,b}-D_{\nu}^{ab}\Theta_{\mu,b}\\
\s_{\mu}^a&\ \to\ \s_{\mu}^a-\Theta_\mu^a
\end{split}
\eqlabel{25}
\end{equation}
which serves to generalize the transformations of \eqref{17}. The vertices arising in \eqref{24} are 
unfortunately not 
consistent with renormalizability.

\section*{Acknowledgments}
Research at Perimeter Institute is
supported by the Government of Canada through Industry Canada and by
the Province of Ontario through the Ministry of Research \&
Innovation. AB gratefully acknowledges further support by an NSERC
Discovery grant and support through the Early Researcher Award
program by the Province of Ontario. Roger Macleod had a useful suggestion.

\end{document}